\documentclass[12pt]{iopart}
\usepackage{graphicx}
\usepackage{pstricks,pst-node}
\usepackage{pst-node,pst-text,pst-3d} 
\usepackage{verbatim}
\usepackage{tabularx}
\relax 
\input epsf.tex 
\def\ii{{\'{\i}}}

\font\af=msbm12
\font\ssmall=cmr10
\def\C{{\af C}}

\def\CM{{\cal M}}

\def\be{\begin{eqnarray}}    
\def\ee{\end{eqnarray}}
\def\Dsl{\,\raise.05ex\hbox{/}\mkern-9.5mu D}
\def\mbox#1#2{\vcenter{\hrule \hbox{\vrule height#2in 
\kern#1in \vrule} \hrule}} 
\def\boxeqn#1{\vcenter{\vbox{  \hrule height2pt \hbox{\vrule
width 2pt \kern3pt\vbox{\kern3pt
\hbox{${\displaystyle #1}$}\kern3pt}\kern3pt\vrule width 2pt}\hrule height2pt}}}

\def\back{{{\raise.4em\hbox{$\, _\backslash\,$}}}}

 2

\font\blackboard=msbm10 \font\blackboards=msbm7
\font\blackboardss=msbm5
\newfam\black
\textfont\black=\blackboard
\scriptfont\black=\blackboards
\scriptscriptfont\black=\blackboardss

\def\frac#1#2{{#1\over #2}}

\def\big R{{\hbox{{\bigfield R}}}}
\def\bbig R{{\hbox{{\bbigfield R}}}}

\font\af=msbm10

\font\afm=msbm12

\def\Z{\hbox{\af Z}}

\def\C{\hbox{\afm C}}

\def\Zm{\hbox{\afm Z}}

\def\I{\hbox{\afm I}}

\mathchardef\imath="717B
\def\inbar{\,\vrule height1.5ex width.4pt depth0pt}
\def\IB{\relax{\rm I\kern-.18em B}}
\def\IC{\relax\hbox{$\inbar\kern-.3em{\rm C}$}}
\def\ID{\relax{\rm I\kern-.18em D}}
\def\IE{\relax{\rm I\kern-.18em E}}
\def\IF{\relax{\rm I\kern-.18em F}}
\def\IG{\relax\hbox{$\inbar\kern-.3em{\rm G}$}}
\def\IH{\relax{\rm I\kern-.18em H}}
\def\II{\relax{\rm I\kern-.18em I}}
\def\IK{\relax{\rm I\kern-.18em K}}
\def\IL{\relax{\rm I\kern-.18em L}}
\def\IM{\relax{\rm I\kern-.18em M}}
\def\IN{\relax{\rm I\kern-.18em N}}
\def\IO{\relax\hbox{$\inbar\kern-.3em{\rm O}$}}
\def\IP{\relax{\rm I\kern-.18em P}}
\def\IQ{\relax\hbox{$\inbar\kern-.3em{\rm Q}$}}
\def\IR{\relax{\rm I\kern-.18em R}}
\font\cmss=cmss10 \font\cmsss=cmss10 at 10truept
\def\IZ{\relax\ifmmode\mathchoice
{\hbox{\cmss Z\kern-.4em Z}}{\hbox{\cmss Z\kern-.4em Z}}
{\lower.9pt\hbox{\cmsss Z\kern-.36em Z}}
{\lower1.2pt\hbox{\cmsss Z\kern-.36em Z}}\else{\cmss Z\kern-.4em Z}\fi}
\def\IGa{\relax\hbox{${\rm I}\kern-.18em\Gamma$}}
\def\IPi{\relax\hbox{${\rm I}\kern-.18em\Pi$}}
\def\ITh{\relax\hbox{$\inbar\kern-.3em\Theta$}}
\def\IOm{\relax\hbox{$\inbar\kern-3.00pt\Omega$}}

\def\CM{{\cal M}}

\def\CH{{\cal H}}

\begin{document}

\title{Vacuum Energy and Renormalization on the Edge}

\author{M. Asorey$^1$, D. Garc\'\i a-\'Alvarez$^2$, J. M.  Mu\~noz-Casta\~neda$^1$}

\address{$^1$ Departamento de F\'\i sica Te\'orica. Facultad de Ciencias.
Universidad de Zaragoza, 50009 Zaragoza. Spain}
\address{$^2$ Department of Physics,  Lancaster University, Lancaster LA1 4YB, United Kingdom}
\ead{asorey@unizar.es}
\begin{abstract}
The vacuum  dependence on  boundary conditions in  quantum field theories
is analysed from a very general viewpoint. From this perspective   
the renormalization prescriptions not only imply the renormalization of the couplings of the theory 
in the bulk but also the appearance of a flow in the space of boundary conditions.
For regular boundaries this flow has a large variety of  fixed points and no cyclic orbit.
The family of fixed points includes Neumann and  Dirichlet  boundary conditions. 
In one-dimensional field theories pseudoperiodic and quasiperiodic boundary conditions 
are also RG fixed points. Under  
these conditions massless bosonic free  field theories are conformally invariant. 
Among  all fixed points only Neumann boundary conditions are infrared stable
fixed points. All other conformal invariant boundary conditions become unstable under some
relevant perturbations.   In finite volumes we analyse the dependence of the vacuum
energy along the trajectories of the renormalization group flow providing an
interesting framework for dark energy evolution. On the contrary,  the renormalization
group flow on the boundary does not affect the leading behaviour of the entanglement
entropy of  the vacuum in one-dimensional conformally invariant bosonic theories.

\end{abstract}

\vspace{0pc}
\hspace{2.3cm} {\it Keywords}: {\ssmall \ Renormalization group, boundary conditions, vacuum energy.}
\maketitle

\section{Introduction}

\hspace{20pt}
The emergence of the dark energy as one of the basic ingredients
of the current standard cosmological scenario, and the absence of
an even vague understanding of its possible origin, opens a window 
to the analysis  of all possible mechanisms that generate  background
energy (see e.g.  \cite{Padmanabhan}  for a review of recent proposals).
The main problem is that the apparent value of the dark energy
is very tiny compared with any physical energy scale. A second problem
is that in a generic quantum field theory there is generation of vacuum
energy and any renormalization prescription  requires a
fine tuning, which is not very convincing without the quantisation of the 
gravitational interaction.

The guess that dark energy 
 might change with the evolution of the Universe 
can be understood even if dark energy is just vacuum energy.
The finite corrections due to finite size of the causal Hubble domain
decrease  as the Universe continues to expand.

The aim of this paper is to analyse the variation of these finite size corrections
under of renormalization group on the space of boundary conditions
for scalar field theories in flat space, although the results are generalisable
for more general backgrounds. 

The dependence of the vacuum energy on the boundary conditions  \cite{deutsch}  is well known since
the discovery of Casimir effect \cite{11} (see \cite{milton,bordag} and \cite{jphys} and references therein for
recent revisions).
 However,   boundaries might also be considered as a source of new, although
peculiar, interactions and therefore can undergo
renormalization \cite{moss, odintsov}. The  renormalization of boundary conditions  
 modifies  the critical behaviour  of the theory \cite{Affleck,Zuber,Cardy}. 
In systems with boundaries or defects, the boundary  RG flow induces a   dynamical behaviour
on  the boundaries.  The dynamics of  D-branes in string theory
emerges in this way \cite{polchinski}.

The renormalization group flow  is analysed from a global viewpoint  in the most general framework 
for boundary conditions  of scalar field theories introduced in Ref. \cite{aim}.
In particular, we consider  the possible existence of  topological transitions \cite{14} induced by 
the renormalization of boundary conditions or
 cyclic orbits in the boundary RG  flow  \cite{wilson}. 
The dependence of the finite size corrections to the vacuum energy and 
vacuum entanglement entropy \cite{sorkin,srednicki} under the boundary RG flow
is analysed from a very general perspective.

\section{Boundary conditions in Field Theory}

\hspace{20pt}
The action which governs the dynamics of scalar field theory in a bounded domain $\Omega$ 
of flat space consists of two different of terms, 
$S(\phi)=S_B(\phi)+S_b(\phi)$.
The first one  
\be 
  S_B(\phi)=\frac12
\int dt \int_\Omega { \sqrt{g}\, d^{D} x} \, \left[|\dot\phi|^2 - |\nabla\phi|^2 -V(|\phi|^2)\right]\,
\ee
is defined in terms of the values of the fields in the bulk. The second  term
\be
S_b(\varphi)=\frac12
\int dt \int_{\partial\Omega} { \sqrt{g_{_{\partial\Omega}}} d^{D-1} x} \left[ |\dot\varphi|^2 \!+\! 
\frac12 \varphi^\ast \partial_n\varphi\! +\!
 \frac12    (\partial_n\varphi^\ast) \varphi\! 
-\! |\nabla\varphi|^2\!\right]  
\label{one}
\ee
depends only on the values of the fields at the boundary $\partial \Omega$  $ ^{\dagger}$
\footnote[0]{$\!\!\!^{\dagger}$
We will assume that the boundary is regular and smooth. See e.g.  \cite{tsutsui}  
 for the peculiarities associated
to the presence of irregular boundaries}.
$g_{_{\partial\Omega}}$ denotes the  metric induced on the boundary by the bulk flat metric,
and  $\partial_n $ is the normal derivative at the boundary
\be
 \varphi=\phi|_{_{\partial\Omega}} \qquad   \partial_n\varphi =\partial_n\phi|_{_{\partial\Omega}}.
\ee
The presence of the boundary term $S_b$ allows  the generation of  local classical equations of 
motion  without requiring any specific type of   boundary conditions \cite{bfsv,saharian}.
Indeed, the gradient term
\be
{\cal V}=\frac{1}{2}\int_\Omega |\nabla \phi |^2
\ee
can be rewritten as
 \be
{\cal V}=\frac{1}{2}\int_\Omega \phi^\dagger  \Delta \phi+\frac{1}{2}\int_{\partial\Omega} \phi^\dagger  \, \partial_n \phi
\label{efive}
\ee
where  $ \Delta$  is the Laplace-Beltrami operator $ \Delta =-{\  \,\partial^\mu}\partial_{\mu} $ .
In the quantum  theory the Laplace-Beltrami operator must  have a real spectrum
in order to have  a selfadjoint Hamiltonian 
\be
\CH=\frac1{2}\sqrt{\Delta+m^2}.
\label{hamilton}
\ee
for the free field theory (The inclusion of interactions does not changes the picture 
\cite{symanzik}).
 This means that the classical fields 
must satisfy boundary conditions which make the operator $\Delta$ selfadjoint. 
The complete set  of boundary conditions which satisfy this requirement \cite{aim}
are   in one-to-one correspondence with the group of unitary
operators of  the boundary Hilbert space $\  L^2(\partial\Omega,\C)$.  For
 any  unitary operator $U\in L^2(\partial\Omega,\C)$, the fields satisfying the 
boundary condition
\be
{\phantom{\Bigl[}  
\varphi- i\, \partial_n\varphi ={   U} \left(\varphi+ i\, \partial_n\varphi
\right)\phantom{\Bigr[} }
\label{const}
\ee
define a domain where $\  \Delta$ is a selfadjoint operator.

In the case of open strings,  the corresponding conformal  1+1 dimensional scalar field theories  is
defined on the space interval  $\Omega =[0,1]\subset \IR$ and there is a large variety of admissible
boundary conditions described by the unitary group $\CM=U(2)$.
The unitary matrices
\be
\hspace{-2cm}
U_D= \begin{pmatrix}
{{-1}&{0}\cr
{0}&{-1}} 
\end{pmatrix}\qquad 
U_N=\begin{pmatrix}{{1}&{0}\cr
{0}&{1}}\end{pmatrix}\qquad 
U_P=\begin{pmatrix}{{0}&{1}\cr
{1}&{0}}\end{pmatrix}
\ee
define  Dirichlet, Neumann 
and  periodic  boundary conditions, which in string theory correspond to
a string  attached to a
D-brane background, free  open  and closed string theories, respectively.

For  higher N-dimensional target spaces,  or N-component strings, 
the set of   boundary conditions becomes $\CM=U(2N)$  which 
includes matrices which interpolate between  one single 
closed string  or  N  disconnected strings \cite{aim}. The topology change is described
in this picture  by a simple change of boundary conditions in $\  L^2(\partial\Omega,\C^N)$
\cite{14}.

If the spectrum of eigenvalues of the unitary operator $U$ does not include
the value $\pm 1$ (i.e.
$\pm 1\notin{\rm Sp}\, U$) the boundary condition (\ref{const}) can be rewritten
as 
\be 
\partial_n \varphi=-i\, {\I\pm U\over \I \mp U}\ \varphi
\ee
 which means that only the boundary values of the fields at the boundary can
 have an arbitrary value  $\varphi $ 
whereas its normal derivative is determined by $U$ and $\varphi $.

The corresponding operator mappings  from unitary into selfadjoint operators 
\be  
A_\pm=-i\, {\I \pm U\over \I\mp U}
\ee
are the celebrated Cayley transforms. The inverse Cayley transform
\be
 U= {\I\mp i A_\pm\over \I \pm iA_\pm}
\ee
recovers the unitary operator $U$ from their selfadjoint Cayley transforms $A_\pm$.

The condition of $\Delta$ being selfadjoint is necessary but not sufficient
to guarantee the unitarity  of the corresponding quantum field theory.
Indeed, in the case of free field theory the Hamiltonian (\ref{hamilton})
must be selfadjoint.
This requires that the spectrum of $\Delta+m^2$ must be not only real but  also
positive which restricts the set of admissible boundary conditions to a subset  $\ \CM$ of
 $\  L^2(\partial\Omega,\C)$. 

Because of the existence of the boundary term in (\ref{efive})
 the Hamiltonian $\CH$ (\ref{hamilton}) is not selfadjoint  if the spectrum of the unitary
operator $U$ intersects the following domain of phase factors  
$$
S^1_{m}=\{{\rm e}^{2\alpha i}; -\pi <\alpha \leq \pi ,0<\alpha<\frac{\pi}{2}- \,\arctan\, {m^2}, {\rm
  or}\, \frac\pi{2}<-\alpha<{\pi}- \,\arctan\, {m^2}\, \}.
$$
In any other case, 
 $-m^2$ is a lower bound
for the spectrum of the operator $\Delta$ and $\CH$ is selfadjoint.
One possible source of unitarity loss is the existence of  edge estates   with large 
negative eigenvalues of operator $\Delta$.

The consistency of the quantum field theory imposes, thus, a very
stringent condition on the type of acceptable boundary conditions, even in the
case of massive theories in order to prevent this type of pathological behaviour of vacuum energy.

For real scalar fields there is a further condition. $U$ has to satisfy
a  CP  symmetry preserving condition
 \be
U^\dagger=U^\ast ,\quad U=U^T .  
\ee
The usual Neumann and Dirichlet boundary conditions
${  U=\pm \I \  }$ satisfy this condition.
In general, for
\be
\  U= \begin{pmatrix}{
{A_1}&{B}\cr
{B^T}&{A_2} }
\end{pmatrix}
\ee
the condition requires that
\be
 A_1=A^T_1,\, A_2=A^T_2 ,\, A_1 B^\ast +B A_2^\dagger =0\ee
\be
 B B^\dagger   +A_1A_1^\dagger =\I, \, A_2A_2^\dagger +B^TB^\ast=\I 
\ee

In particular, the quasi-periodic condition
${\varphi}(L)= M^{-1}{\varphi}(0),\ 
 \partial_n{\varphi}(L)= M \partial_n{\varphi}(0)$ is also compatible if $ M=M^t=M^\ast$.

In the case of one single  real massless scalar the set of compatible
boundary conditions  has  two connected components: $\CM_0$
 given by the operators of the form 
\be U_{\beta}=\cos\beta\,{ \  \I} + i \sin \beta\,
  {\  \sigma_y},
\nonumber
\ee
and $\CM_1$ given by 
\be U_{\alpha}=\cos\alpha\,{ \  \sigma_z} + \sin \alpha\,
  {\  \sigma_x}\, .
\ee
 $\CM_0$  includes Neumann ($\beta=0$)  and Dirichlet ($\beta=\pi$) conditions; and $\CM_1$
contains the  quasi-periodic boundary conditions 
\be
\varphi(L)={\tan \frac\alpha{2}}\,\varphi(0);\qquad \partial_n{\varphi}(L)=\left({\tan \frac\alpha{2}}\right)^{-1}{\,}\partial_n{\varphi}(0)
\label{quasi}
\ee
which include periodic ($\alpha=\frac\pi{2}$)  and antiperiodic ($\alpha=-\frac{\pi}{2}$)  boundary
conditions.

\section{Boundary Conditions and Renormalization Group}

\hspace{20pt}
Since boundary conditions appear more naturally in the Schr\"odinger picture of
field theory and the  theory is plagued  of ultraviolet singularities
some doubts were raised about their relevance for the quantum
field theory. The pioneer work of Symanzik \cite{symanzik} confirmed
the consistence of the standard picture even in presence of bulk renormalizable
interactions (see  \cite{elizalde} for an explanation of a recent controversy \cite{jaffe}).

Moreover, there is a renormalization of the very boundary conditions because the  
boundary terms are the source of new interactions.

The renormalization group can be defined in the continuum approach by
\be
\phi_{\Lambda}\left(\frac{x}{\Lambda}\right)=\Lambda^{[\phi]} [\phi(x)-\xi_\Lambda(x)]
\ee
by means of a fluctuating field 
$\xi_\Lambda$ with short range fluctuations of order $ \frac{1}{\Lambda}$.
This implies that the  boundary condition
\be
\partial_n\varphi=\mathrm{A}\, \varphi
\ee
is renormalised  to
\be
 \partial_n \varphi_{\Lambda}= \mathrm{A_{\Lambda}} \varphi_{\Lambda},
\ee
since 
\be
\!\!\!\!\! \partial_n\phi_{\Lambda}\left(\frac{x_b}{\Lambda}\right)=\Lambda^{[\phi]+1}[\partial_n\phi(x_b)-\partial_n\xi_\Lambda(x_b)]
=\mathrm{A}\, \Lambda^{[\phi]+1}\phi(x_b)=\mathrm{A}_\Lambda \phi_\Lambda\left(\frac{x_b}{\Lambda}\right)
\ee
with  $A_\Lambda=\Lambda A$.
For more general boundary conditions the continuum renormalization group
is given by
\be
 \Lambda\, U_\Lambda^\dagger \partial_{\Lambda} U_\Lambda= \frac12 \left(U^\dagger_\Lambda -U_\Lambda\right)
\ee
or
\be
{\phantom{\Bigl[}   U_t^\dagger \partial_{t} U_t= \frac12 \left(U^\dagger_t -U_t\right){\phantom{\Bigl]}}}
\label{RGf}
\ee
for  $ \Lambda=\Lambda_0\, {\mathrm{e}}^t.$
Fixed points correspond, therefore,  to self-adjoint boundary conditions $ U^\dagger=U$.
In particular,
{ Dirichlet  and Neumann}  ($ U=\mp$ \I) are renormalization group fixed points.

For mixed boundary conditions the RG flows from   Dirichlet ({UV}) toward 
 Neumann ({IR}) conditions.
\be
 U=\mathrm{e^{2i \arctan e^{-t}}} {\mathrm {\I}}.
\label{flip}
\ee

Critical exponents can be identified with the eigenvalues of the  matrix $U_c$ at the
 fixed points. Since $U_c$  is also hermitian   all  critical exponents are either $1$ or $-1$
and  there is no room for cyclic orbits.
It is well known, however,  that  some quantum systems with singular 
boundaries and singular interactions \cite{wilson,aes} 
exhibit cyclic renormalization group flows. Moreover, some topological field theories (e.g. Russian doll models) 
present a similar behaviour \cite{sierra}. In scalar field theories, this phenomenon simply does not occur for 
regular boundaries.   For the same reasons topological transitions do not occur for finite scale transformations since
the flip  of eigenvalues from  $ -1$ to $+1$   requires a change in the parameter $t$ of
the flow from $-\infty$ to $\infty$  as  in (\ref{flip})).

\section{Conformal Invariance and boundary conditions}

\hspace{20pt}
In 1+1 dimensions   the  theory of massless scalar fields  is formally conformal
invariant. However,  boundary conditions might break this symmetry
 \cite{Affleck,Zuber,Cardy}.

Conformal invariance is only preserved  if the boundary conditions are stable under 
 the boundary renormalization group flow.
The fixed points can easily be  identified. 
For a complex scalar field, besides the above mentioned  fixed points, which correspond to
 Dirichlet, Neumann and pseudo-periodic boundary conditions and obviously are
conformal invariant, there are fixed points corresponding to quasi-periodic boundary
conditions (\ref{quasi}).
They also preserve the conformal symmetry.  

In 1+1 dimensions this exhausts
the whole set of conformal invariant boundary conditions. 
Any other boundary condition flows toward one of these fixed points.
The most stable fixed point corresponds to Neumann  conditions because
all its critical exponents are $+1$. The most unstable is that of   Dirichlet 
conditions since  all critical exponents $-1$. This is compatible with the fact that 
the neighbourhood of Dirichlet boundary conditions is plagued of singularities 

 Periodic, quasi-periodic and pseudo-periodic fixed points present relevant and irrelevant
perturbations with critical exponents $\pm 1$, respectively. Negative values 
label the possible instabilities. Implications of these results
for  string theory are well known. 
Periodic boundary conditions, appear as  attractors of systems with
quasi-periodic  and pseudo-periodic conditions which stresses 
the stability of closed string theory vacuum.
For open strings the  (stable) attractor points are  standard free strings
(Neumann). Any other  boundary 
condition flow toward one of those fixed points.

Notice that the absence of topological transitions in the boundary renormalization group 
flow is a consequence of the fact that all relevant perturbations are always 
associated with  -1  critical exponents.

In higher dimensions  ($D>1$)  conformal invariance requires, even in the massless case $m=0$,
that Neumann boundary conditions have to be  modified 
 in order to preserve conformal invariance with a term
\be
\partial_n{\varphi} =\frac{D-1}{4\, D}K\, \varphi,
\ee
proportional to  the extrinsic curvature  $K$  of the boundary.

In the case of singular boundaries some more interesting boundary renormalization group flows 
arise  (see e.g. \cite{tsutsui} for a review):
fixed points and cyclic orbits of the boundary renormalization group flow 
can appear \cite{wilson,sierra,aes} and 
conformal invariance  can be partially  broken to a discrete subgroup
\Zm \,  \cite{aes}.

\section{Vacuum energy and boundary conditions}

\hspace{20pt}
The infrared properties of quantum field theory are very sensitive to boundary
conditions \cite{karpacz}. In particular,  physical properties of the quantum 
vacuum state like  the vacuum energy may exhibit  a very strong dependence on 
the type of boundary conditions.  This can be explicitly shown in the simple case of 
 a  massless field  defined on a finite one-dimensional interval $[0,L]$.

For  pseudoperiodic boundary conditions defined by the unitary operator
\be
  U_\theta =\cos  \theta\, {\  \sigma_x} - \sin \theta \, \  \sigma_y: \quad  \varphi(L)={\rm e}^{i \theta} \varphi(0)
\label{pseudo}
\ee
the Casimir vacuum energy (see e.g.  Ref. \cite{bordag} and references
therein) is given by
\be  E_0=\frac{\pi}{
  L}\left(\frac{1}{12}-{\min_{n\in\Z}}\left({ \frac{\theta}{2\pi}\,}{+n}-\frac1{2}\right)^2\right)
\ee

The vacuum energy dependence on $\theta$ is in this case  relatively smooth. The only 
 cuspidal point at $\theta=0$  corresponds to periodic
boundary conditions. A completely regular  behaviour is obtained
for Robin boundary conditions
\be
  U={\rm e}^{2 \alpha i}\I :  \, \quad  \partial_n{\varphi}(0)=\tan \alpha{}\, \varphi(0),\
\partial_n{\varphi}(L)={\tan}\, \alpha{}\, \varphi(L) 
\label{robs}
\ee
which smoothly  interpolate between Dirichlet ($\alpha=\frac{\pi}{2}$) and Neumann ($\alpha
=\pi$)  conditions when $\alpha$ is restricted to  the interval  $\alpha\in
[\frac\pi{2},\pi]$ \cite{aa, cavalcanti, farina} .

Finally,  the   Casimir energy for quasi-periodic boundary conditions  \cite{gift}
\be
 E_0=\frac{\pi}{
  L}\left(\frac{1}{12}-{\min_{n\in\Z}}\left(\frac{  \alpha}{2\pi}+n+\frac1{4}\right)^2\right)
\label{qp}
\ee
is also dependent on the choice of the parameter $\alpha$.
Two particularly  interesting cases  are
given by $\alpha=0$, 
$\  U_Z={\  \sigma_z} ; \quad \varphi(L)=0,\ \partial_n{\varphi}(0)=0 $
and  $\alpha=\pi$, 
$\  U'_Z={\  \sigma_z} ; \quad \varphi(0)=0,\ \partial_n{\varphi}(L)=0 
$
 which correspond to  a  Zaremba (mixed) boundary conditions: one boundary is
Dirichlet and the other Neumann.

\section{  Vacuum entanglement entropy}

\hspace{20pt}
The dependence of vacuum energy  on   boundary conditions
seems to suggest that many other observables may suffer the same effect.
In particular, one may wonder  whether or not  the entropy of the system is 
 dependent on the type of  conditions
that constrain the values of the fields  at the boundary.
The entropy of the field theory at finite temperature scales with the volume
of physical space. Only in quantum gravity or string theory the entropy can scale with
the area of black hole  horizon. However, in field theory it is possible
\medskip

\centerline
{  \epsfbox{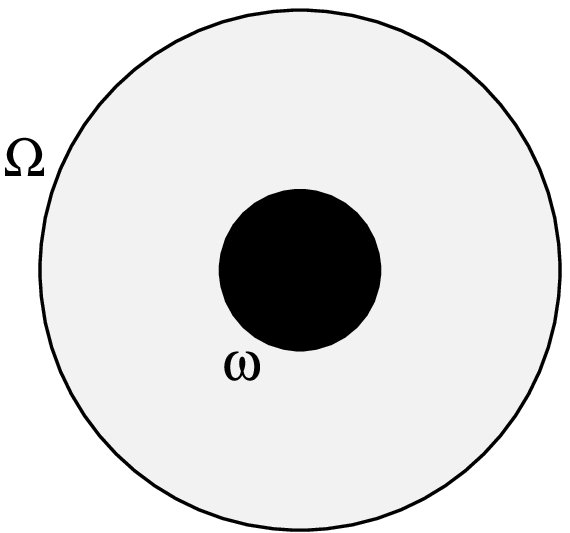}}

\bigskip

\centerline{\small {\bf Figure 1.}{ Information loss  by integration over
the fluctuations of the fields inside the domain $\omega$ }}

\bigskip
\noindent
to generate a mixed state from the pure vacuum state $\Psi_0$  by integrating out
the fluctuating modes in a bounded domain $\omega$ of the physical space
$\Omega$ (see Figure 1)
\be
{\rho_\omega}=\int_{\omega}  \Psi_0^\ast \Psi_0.
\ee
 The entropy 
of this  state  $ S_\omega= -Tr \,\rho_\omega\, \log \rho_\omega$, although  ultravioletly 
divergent, provides a  measure of  the degree of entanglement of the vacuum state.  In the case of 
 a free massless real scalar field theory in  one-dimensional spaces ($D=1$) 
this entropy  scales logarithmically  
with the size $l_\omega$ of $\omega$
and the  ultraviolet cut-off $\epsilon$
 introduced to split apart  the domain $\omega$ and its complement 
$\Omega\backslash\omega$
\be
 S_\omega=\frac1{3} \log\frac{ l_\omega} { \epsilon},
\label{eone}
\ee
and in $D=2$ dimensions it scales linearly with the perimeter $R_\omega$ of $\omega$ 
\be
 S_\omega=c_2\, \frac{ R_\omega}{\epsilon}\, - \, { \gamma}
\label{etwo}
\ee
and in $D>2$ dimensions  as the volume of the boundary of $\omega$
\be
 S_\omega=c_{_D} { V_\omega} \epsilon^{1-D}.
\label{ethree}
\ee
In particular in  three-dimensional spaces it scales  with  the area   of the boundary of $\omega$
like  in the presence of a blackhole \cite{sorkin,srednicki}.
Although the coefficients of the leading terms $c_2,c_{_D}$ in (\ref{etwo}) and (\ref{ethree}) have been explicitly computed,
they are not universal because  they obviously  depend on the choice of the UV cutoff $\epsilon$. On the
contrary,   the coefficient $c_1=1/3$ 
of the logarithmic term in (\ref{eone}) is universal and does coincide with one third of the central charge of 
the corresponding conformal field theory. Similarly,  the finite  $\gamma$  term in (\ref{etwo})
 is also universal in $D=2$ dimension
and is related to a degree of topological entanglement \cite{kitaev}.

It is remarkable that in $D=1$ the coefficient $c_1=1/3$ is also independent of the choice of boundary condition
in $\Omega$. This in contrast with what happens  for the finite size corrections to vacuum energy.
The coefficient  of the $1/L$ term is also proportional to the central charge but in that case
the corresponding factor is very sensitive to the type of boundary conditions imposed at the boundary
of $\Omega$. The above results  indicate that  whereas the  
Casimir energy is closely related with the infrared properties
of the conformal theory which are sensitive to the boundary conditions, the entanglement
entropy is rather associated to the  behaviour at the interface between $\omega$ and its complement
$\Omega\backslash\omega$ which do not depend on the choice of boundary conditions at the edge 
of the physical space.

\section{Conclusions}

\hspace{20pt}
The description of regular boundary conditions in terms of unitary matrices provides a very useful framework for
the description of the boundary renormalization group flow and the breaking of conformal invariance due
to boundary effects. Neumann  conditions turn out to be the only  boundary 
conditions which are absolutely stable under RG  flow. 
All other boundary conditions may have  some relevant perturbations which are
the source of RG instabilities. However, the global structure of the flow does not permit  topological transitions.

The finite size corrections to vacuum energy are very sensitive to the choice of boundary conditions which discriminate
between the different fixed points of the renormalization group flow. On the contrary, the leading contribution to
 entanglement entropy of the vacuum  is insensitive, for one-dimensional massless scalar field theories,
to the change of boundary conditions. In D=2 dimensions  the same property holds for the finite
 correction to the entanglement entropy of massless scalar theories. This fact, is very relevant for 
the implementation of quantum  codes with topological stability \cite{kitaev}.  However, these properties do not
hold for the leading terms contributing to the entanglement entropy.

\section*{Acknowledgements}
\hspace{20pt}
We thank E. Elizalde,
J.G. Esteve, S. Odintsov and G. Sierra
for interesting discussions on closely related subjects. 
This work is partially supported by CICYT (grant FPA2004-02948)
and DGIID-DGA (grant2006-E24/2).

\bigskip

\end{document}